\documentclass[sigplan,10pt,a4, nonacm, anonymous=false]{acmart}
\renewcommand\footnotetextcopyrightpermission[1]{}
\AtBeginDocument{%
  }

\usepackage{subcaption}
\usepackage{tcolorbox}
\usepackage{csquotes}
\usepackage{xspace}

\newcommand\s[1]{(\S\ref{#1})\xspace}
\newcommand\one{(\emph{i})\xspace}
\newcommand\two{(\emph{ii})\xspace}

\newcommand\newverbaux[1]{#1}%
\DeclareRobustCommand\newverb[1]{%
  \begingroup\catcode`\^^I=12 \innernewverb{#1}%
}%
\begingroup
\makeatletter \catcode`\Z=14 \catcode`\%=12
\newverbaux{Z
  \endgroup
  \NewDocumentCommand\innernewverb{m+v}{Z
    \endgroup
    \@ifdefinable{#1}{Z
      \DeclareRobustCommand#1{\begingroup\newlinechar=\endlinechar\scantokens{\endgroup#2%}}Z
        }Z
      }Z
    }%

\newverb\cgroup{\verb|cgroup|\xspace}
\newverb\cgroups{\verb|cgroup|s\xspace}
\newverb\gse{\verb|gse|\xspace}
\newverb\gses{\verb|gse|s\xspace}
\newverb\se{\verb|se|\xspace}
\newverb\cfsrq{\verb|cfs_rq|\xspace}
\newverb\cfsrqs{\verb|cfs_rq|s\xspace}
\newverb\vruntime{\verb|vruntime|\xspace}
\newverb\customrq{\verb|rq|s\xspace}

\newcommand\note[2]{{\color{#1}\bf #2}}
\newcommand\mort[1]{{\note{red}{mort: #1}}}
\newcommand\morti[1]{{\note{red}{mort: \begin{itemize}#1\end{itemize}}}}
\newcommand\ek[1]{{\note{blue}{EK: #1}}}

\newcommand\cfsllf{CFS-LAGS\xspace}

\begin{document}


\title{Mitigating context switching in densely packed Linux clusters with Latency-Aware Group Scheduling}

\author{Al-Amjad Tawfiq Isstaif}
\affiliation{
  \institution{University of Cambridge}
  \city{Cambridge}
  \country{UK}
}
\email{aati2@cam.ac.uk}

\author{Evangelia Kalyviannaki}
\affiliation{
  \institution{University of Cambridge}
  \city{Cambridge}
  \country{UK}
}
\email{ek264@cam.ac.uk}

\author{Richard Mortier}
\affiliation{
  \institution{University of Cambridge}
  \city{Cambridge}
  \country{UK}
}
\email{rmm1002@cam.ac.uk}

\begin{abstract}
  Cluster orchestrators such as Kubernetes depend on accurate estimates of node capacity and job requirements. Inaccuracies in either lead to poor placement decisions and degraded cluster performance. In this paper, we show that in densely packed workloads, such as serverless applications, CPU context switching overheads can become so significant that a node's performance is severely degraded, even when the orchestrator placement is theoretically sound. In practice this issue is typically mitigated by over-provisioning the cluster, leading to wasted resources.

  We show that these context switching overhead arise from both an increase in the average cost of an individual context switch and a higher rate of context switching, which together amplify overhead multiplicatively when managing large numbers of concurrent \emph{cgroups}, Linux’s group scheduling mechanism for managing multi-threaded colocated workloads. We propose and evaluate modifications to the standard Linux kernel scheduler that mitigate these effects, achieving the same effective performance with a 28\% smaller cluster size. The key insight behind our approach is to prioritise task completion over low-level per-task fairness, enabling the scheduler to drain contended CPU run queues more rapidly and thereby reduce time spent on context switching.
\end{abstract}

\settopmatter{printfolios=true}
\maketitle

\section{Introduction}

In \emph{serverless computing} users are charged based on the resource usage of
individual \emph{functions}, typically implemented as containers hosted on
virtual machines (VMs). The default Linux Completely Fair Scheduler (CFS)~\cite{CFS} is
widely used in container-based serverless systems~\cite{opensource-serverless,RunD,Firecracker} to implement group scheduling
\cite{cfs-group}. This enhances CFS' weighted fair scheduling by treating all
threads of a process as a single entity to prevent processes gaining additional
CPU time by simply creating more threads. Control Groups~(\emph{cgroups}) provide a user space interface for group scheduling, imposing
a hierarchical structure on processes and threads, distinguishing between
workloads with different priorities~\cite{resctl, RunD}~\s{sec:background}.

The serverless model shifts the cost of idle VMs from users to the service
provider managing the underlying server capacity, making it ideal for
intermittent and bursty workloads~\cite{Shahrad,HuaweiTraces}. Additionally, container-based serverless
platforms~\cite{FaaSNet, Brooker,FaaSNet,Owl,RunD} often keep idle functions
active to avoid the costs of container initialisation. This can result in a very
large number of containers that consume a minimum amount of resource as far as
the cluster scheduler is concerned but in practice are largely idle, resulting
in an under-utilised---equivalently, over-provisioned cluster~\cite{XFaaS,HuaweiTraces,Shahrad}.

The natural way to reduce this under-utilisation is to increase the number of
functions co-located on each worker node~\cite{Jiango,Gsight,Golgi,Owl}. This
approach can result with a higher-density deployment due to the small memory
footprint of most functions. For example, a significant portion of AWS Lambda
functions operate with a minimum of 128MB memory, and most applications seldom
exceed 400MB~\cite{Shahrad}. Li et al~\cite{RunD} report hosting over 2,500
containers of 128MB each on a 384GB node in Alibaba's serverless platform.
On modern multicore machines, when there are relatively large numbers of functions at the same priority level, we observe that CFS’s group scheduling structure significantly increases the time spent context switching, reaching 5---20\% CPU utilisation. This is due to increases in both the average cost of a single context switch and the rate of context switching, which combine multiplicatively~\s{sec:motivation}.

Our CFS-Latency-Aware Group Scheduling (\cfsllf) is a variant of Linux's standard CFS scheduler that mitigates these effects by enabling task completion for functions through two mechanisms: \one relaxing the strict fairness criteria, and \two ensuring available  CPU cores are used to maximum effect. As such, \cfsllf allows functions to exit the system more quickly, thereby shortening run queues and mitigating their associated overheads.
This requires non-trivial modifications to CFS’s highly concurrent codebase due to its per-cgroup, per-core run queue structure~\cite{WastedCores,Ipanema}~\s{sec:cfs-llf}.

We show that \cfsllf significantly improves server throughput under moderate to high demand, without detriment under low demand. Furthermore, \cfsllf limits the growth of scheduling overhead, enabling greater function colocation due to more robust handling of overload. This increases average server utilisation by 10\%, enabling consolidation of functions onto fewer nodes in a cluster and reduction in cluster size of 28\%~\s{sec:cfs-llf-evaluation}.
Our contributions are:
\begin{itemize}
\item a quantitative analysis of the causes of unreliable scheduling behaviour in serverless clusters~\s{sec:motivation};
\item the design and implementation of the new \cfsllf scheduler, which mitigates scheduling overhead effects and makes cluster worker nodes more resilient under overload in serverless workloads~\s{sec:cfs-llf}; and
\item a thorough evaluation of \cfsllf using both synthetic workloads and real-world traces, demonstrating that it significantly reduces scheduling overheads while improving latency and cluster utilisation~\s{sec:cfs-llf-evaluation}.
\end{itemize}

\section{Linux scheduling for serverless clusters}
\label{sec:background}

This section outlines the critical role of Linux group scheduling~\cite{cfs-group} via CFS in managing CPU allocation within densely packed serverless clusters. The Linux kernel is a general purpose operating system supporting an
enormous range of workloads and schedulers. The Completely Fair Scheduler
(CFS)~\cite{CFS} is the default Linux scheduler. CFS is widely used for
lightweight VMs~\cite{RunD, Firecracker} and container-based serverless
systems~\cite{opensource-serverless, Kaffes2021PracticalSF}. CFS treats both
processes and kernel threads as \emph{tasks}, aiming to fairly share CPU among all tasks.

CFS can run alongside other Linux scheduling policies, which are organised into distinct scheduling classes. Each class corresponds to a concrete scheduler implementation that may support multiple policies. All scheduling classes are managed by a meta-scheduler, also known as the core scheduler, which executes them in order of priority, beginning with the real-time class. These policies enable the colocation of diverse workload types — for example, latency-sensitive workloads benefit from placement in the real-time scheduling class, while best-effort workloads may be assigned to the \verb|SCHED_IDLE| policy~\cite{cfs-schedidle}. While these various policies and classes are useful for colocating workloads with differing priorities~\cite{ModernBVT}, our focus is on the colocation of functions with similar priorities in serverless environments, where CFS provides the most robust option for sharing CPU resources.

\subsection{CFS scheduler}\label{sec:cfs-background}

CFS shares CPU cores among tasks via an approximate processor-sharing
scheduling policy~\cite{Hermod} where time is proportionally allocated to
each task, weighted by task priority. CFS is also work-conserving~\cite{WastedCores,Ipanema}, ensuring no CPU core remains idle if
tasks are waiting in other cores' run queues. Where strict CPU usage limits are needed
to provide isolation between tasks, CFS bandwidth control~\cite{cfs-quota}
throttles tasks that exceed their allocated CPU time. However, this effectively
limits the use of work-conserving scheduling and so this behaviour is disabled
in high-density deployments that require high levels of task colocation on CPUs~\cite{k8s-dont-use-limits}.
These principles are akin to those used in widely deployed multi-core schedulers like FreeBSD's
ULE~\cite{OSSchedulers} and Xen's Credit scheduler~\cite{Xen-CPU}. In practice, each task—whether a process or a thread—is represented under CFS as a scheduling entity~(\se).

Linux supports grouping tasks into \emph{control groups} (\emph{cgroups}),
collections of tasks that are managed as a unit,
sharing resource quotas and other metrics.
CFS implements group
scheduling~\cite{cfs-group} in support of cgroups by representing them as
\emph{group scheduling entities} (\gse), each of which corresponds to a \cgroup
on a specific CPU core and is used to track metrics for all tasks within that
\cgroup. Each \gse is placed on a run queue (\cfsrq) owned by its parent \gse.
The result is a tree of \cfsrqs per core that reflects the \cgroup hierarchy
ultimately rooted at the top-level \cfsrq for that core~\cite{resctl,RunD}. Each
\cfsrq is internally implemented as a red-black tree with leaves corresponding
to either group or normal scheduling entities (\gse or \se). As well as managing
scheduling within individual \cfsrqs, the kernel also performs load balancing
across the queues of different cores.

The underlying scheduling metric used by CFS is \emph{virtual runtime}
(\vruntime), tracked per task in the \emph{scheduling entity}~(\se) or per
group of tasks in the corresponding group entity~(\gse). CFS provides fairness
by ensuring every runnable task receives a round-robin CPU allocation within
each \emph{scheduling period}. Further, CFS grows the period as needed to ensure the CPU
allocation per-task is above a minimum threshold (\verb|min_granularity|). CFS implements this by recording the
runtime spent by each \se in an attribute \verb|se->vruntime|, and prioritising
entities with the lowest runtime so that no scheduling entity is starved. When a
tasks uses all its allocation, it is preempted and its \vruntime updated before
it is inserted back into the red-black tree implementing the \cfsrq.
A task that is preempted must
wait until its \vruntime is once again the minimum value in all its descendant \cfsrq run queues before it
may be re-scheduled on a core.

Recent implementations of CFS (\texttt{kernel/sched/fair.c}) adopt a new policy to realise fairness~\cite{EEVDF-linux}, inspired by Earliest Eligible Virtual Deadline First (EEVDF)~\cite{EEVDF}.
Under this policy, users may
explicitly configure a task's time slice with more accurate accounting for this
slice through a ``lag'' metric. Positive lag means the task needs more CPU time,
while negative lag means it has had more than its share. A task becomes runnable
whenever its lag is positive and this is used to calculate a virtual deadline by
adding its time slice to the time it became runnable. This virtual deadline is used to order tasks in the existing CFS run-queue structure (i.e.,~the per-cgroup, per-core \verb|cfs_rq|s), replacing the original virtual runtime.
In this way, tasks with a shorter time slice will have an earlier deadline which has
shown to be helpful for latency-sensitive tasks without the risk of overtaking
the system with real-time policies~\cite{EEVDF-slice}.

\subsection{Scheduling container-based functions}

We next summarise previous efforts to increase packing efficiency and enhance utilisation in serverless clusters. There is a range of container-based serverless systems all of which are
Linux-based; Li et al~\cite{opensource-serverless} surveys several serverless
frameworks based on Kubernetes~\cite{k8s-scheduler}, perhaps the most common
cluster orchestration framework. Widely used examples of such frameworks include
Knative~\cite{Knative} and OpenFaaS~\cite{OpenFaaS}.

\begin{figure}[h]
  \centering
  \includegraphics[width=\linewidth]{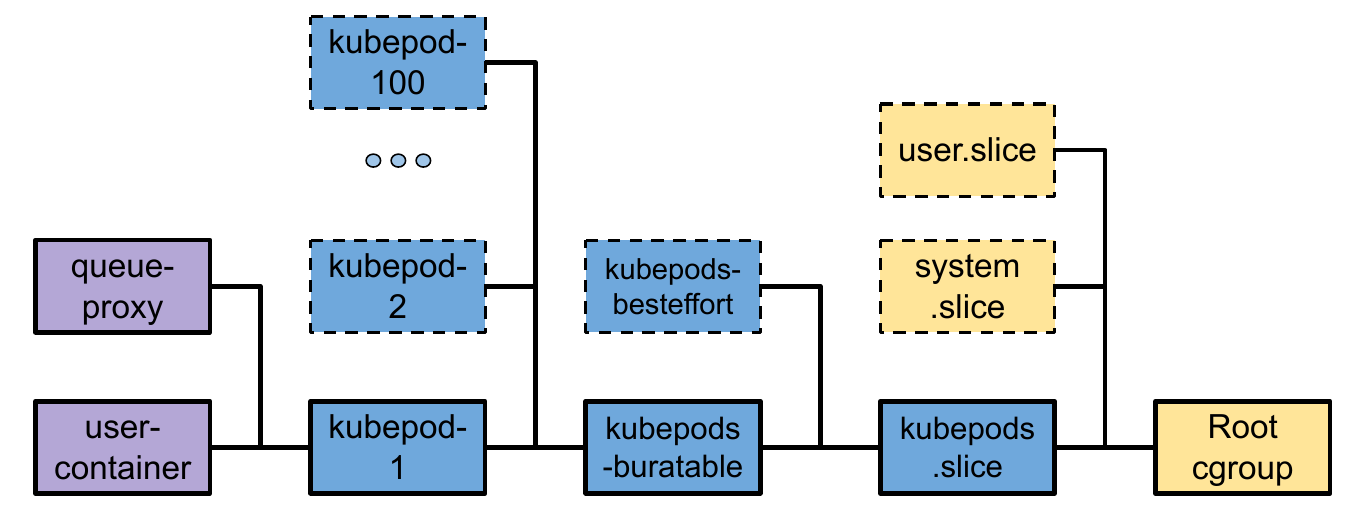}
  \caption{\label{fig:knative-cgroups} The \emph{cgroup} hierarchy instantiated
    for 100 Knative functions on a Linux worker node in a Kubernetes cluster.
    Dashed boxes have
    child entries elided for space, while solid boxes have all child entries
    shown.}
\end{figure}

Deploying a number of Knative functions on a cluster node results in the
cgroup structure depicted in Figure~\ref{fig:knative-cgroups}. Knative
implements the typical sidecar pattern in which a user function is deployed as a
\verb|user-container| and proxied by a \verb|queue-proxy|. The sidecar approach
helps provide generic observability and security features for functions out of the
box. The two containers are deployed as a single Kubernetes \emph{pod}, the
smallest deployment unit in a Kubernetes cluster. By default, all Kubernetes
pods are placed in the \verb|kubepods-burstable| \cgroup, while the
\verb|kubepods-besteffort| entry is reserved for pods which have no resource
configuration. On a multi-core machine, the net result can be several hundred \cfsrq instances over which
CFS must iterate to ensure fair CPU slice allocation.

The key feature that distinguishes serverless from conventional container
workloads is \emph{scale-to-zero} whereby idle containers are removed from the
cluster after a period of inactivity (the \emph{keep-alive period}).
Correspondingly, the ingress controller or API gateway that receives incoming
function invocations must be able to route them to a running container,
initialising a new one if necessary. This creates a particular challenge for
serverless workloads: the cost and overhead associated with the extended
cold-start time of functions. Such cold-starts have been shown to significantly
increase response latency, e.g.,~to 50---85\,s in Alibaba's
platform~\cite{FaaSNet}.

The keep-alive period seeks to reduce cold-starts by delaying the scale-to-zero
behaviour, keeping a container running for a period while idle. Analysis of the
Azure Function Trace data by Li et al~\cite{Golgi} shows that 91.7\% of
functions are invoked once per minute or less, with keep-alive periods typically
set at several minutes. As 96\% of function executions complete within 60\,s
such keep-alive policies result in resource under-utilisation. If containers
are also configured to process multiple concurrent requests, they will need to
be allocated sufficient resource to handle the peak concurrency, leading to
further under-utilisation most of the time.

In response, providers increase resource utilisation by overcommitting
cluster resources, shown to be promising for serverless workloads~\cite{Owl,
  Golgi, Iluvatar, RunD, FirePlace}. By statistically multiplexing workloads,
under the assumption that multiple independent workloads are unlikely to reach
peak resource usage simultaneously, providers can reduce server costs by
increasing workload colocation rather than over-allocating resources per
function. To do so, cluster administrators must activate an overcommitment
policy~\cite{TakeItToTheLimit, Owl,Golgi} that deliberately allocates a task
fewer resources than requested.

Such approaches~\cite{Golgi,Gsight,Jiango} increase workload colocation and reduce contention. However, they do not explore the impact
of kernel scheduling on worker nodes, which we show plays a critical
role during high utilisation. Rather, they focus on improving utilisation
to a point that does not significantly impact the tail latency (i.e.~95th
percentile) under the default Linux CPU scheduling setup. For example, authors of
Gsight~\cite{Gsight} report an average CPU utilisation around 60\% and much
lower utilisation of memory resources with an average around 30\%. The
production study at Meta~\cite{XFaaS} shows that tolerating host inefficiency
can bring significant reduction in server costs, achieving 66\% average daily
utilisation at the expense of \textbf{80$\times$ increase in task tail latency}
compared to median latency of 1\,second.

\section{Quantifying the scheduling overheads}
\label{sec:motivation}

Serverless workloads are challenging to study as, in reality, they involve a
large number of functions being allocated to nodes. This gives rise to $O(n^k)$
possible placements where $n$ is the number of functions and $k$ the number of
functions allowed on a given node~\cite{Gsight,Jiango}. We therefore introduce a
microbenchmark to systematically explore the effects of increasing number of
functions colocated on a node. To the best of our knowledge there is no such
microbenchmark available. We do so under the pessimistic assumption that
workload peaks for different functions may overlap on a given node. Our premise is that although
cluster schedulers aim to minimise the overlap of such peaks~\cite{Gsight, Jiango,Golgi}, it is
difficult to predict future usage patterns and so such overlaps are
inevitable~\cite{TakeItToTheLimit, Christofidi}.

\begin{figure}
  \centering
  \includegraphics[width=1\linewidth]{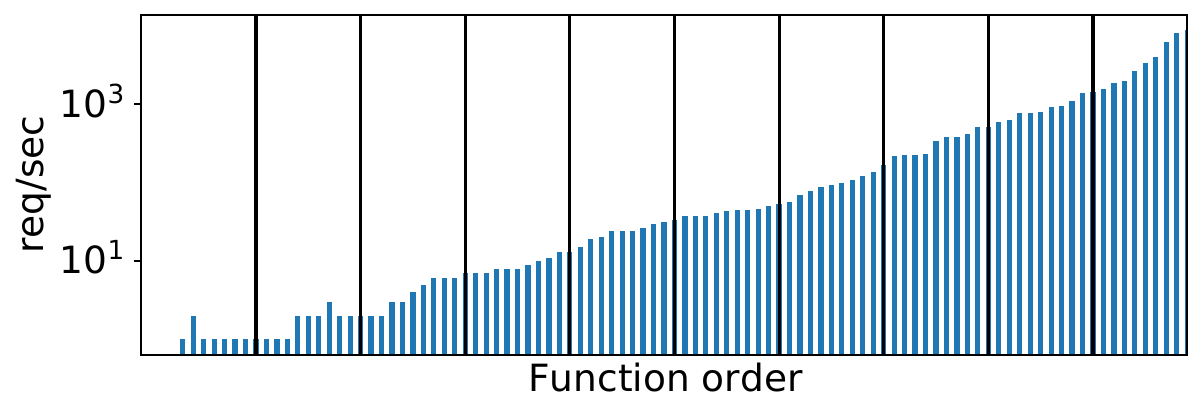}
  \caption{\label{fig:skewed-demand}Sorted per-function distribution of
    requests/second across 5\,minute trace segments in the Azure Function
    Invocation Trace~\cite{AzureFunctionsInvocation}, separated by vertical
    black lines into 10 \emph{demand bands}.
  }
\end{figure}

We start from a real-world function invocation trace, the Azure Functions
Invocation Trace~\cite{AzureFunctionsInvocation}. This dataset includes precise
inter-arrival times for 119 functions collected from the Azure serverless
platform over a 14-day period. We are concerned with the impact of contention
during short, 5\,minute intervals during which the kernel scheduler plays a
critical role. We thus segment each function trace into 5-minute intervals, and
then extract those segments with the largest number of invocations.

Figure~\ref{fig:skewed-demand} depicts the resulting distribution, sorted by
load and divided into ten equal sized \emph{demand bands}. The distribution is
heavily skewed: peak demand over the two-week trace for most functions
reaches only tens or low hundreds of requests/second, but rises to thousands for
the busiest functions. We then synthesise traces with a range of workload
colocation levels from these trace segments to reflect varying deployment
density by drawing equally from each band. The number of functions allocated to
a node is scaled as a multiple of the number of CPU cores, e.g.,~12, 24, 36,
giving an increasing \textit{density factor} of 1$\times$, 2$\times$, 3$\times$.

Scheduling overhead is quantified as the total time spent on context switching,
managed by the \textbf{Linux core scheduler}’s \verb|schedule()|
function.\footnote{\url{https://elixir.bootlin.com/linux/v5.18.19/source/kernel/sched/core.c\#L6497}}
This measurement is obtained using \verb|ftrace| function profiling. We validate that the instrumentation does not significantly alter system behavior by comparing results collected with and without profiling enabled, ensuring that there is no statistically significant difference in latency. \verb|ftrace|
also enables collection of further trace data including the number of context
switches and the average execution time of the \verb|schedule()| function, as
well as a detailed trace of functions executed within the kernel.

We detail the evaluation setup in \S\ref{sec:cfs-llf-evaluation}, confirming that the problem persists in the latest stable EEVDF release (v6.12). However, we report results on CFS v5.18, the base of our \cfsllf~implementation (\S\ref{sec:cfs-llf-implementation}), ensuring that performance differences are attributable to the introduced changes.

\subsection{Overhead on a stand-alone host}\label{sec:motivation:host}

\begin{figure}
  \centering
  \includegraphics[width=1\linewidth]{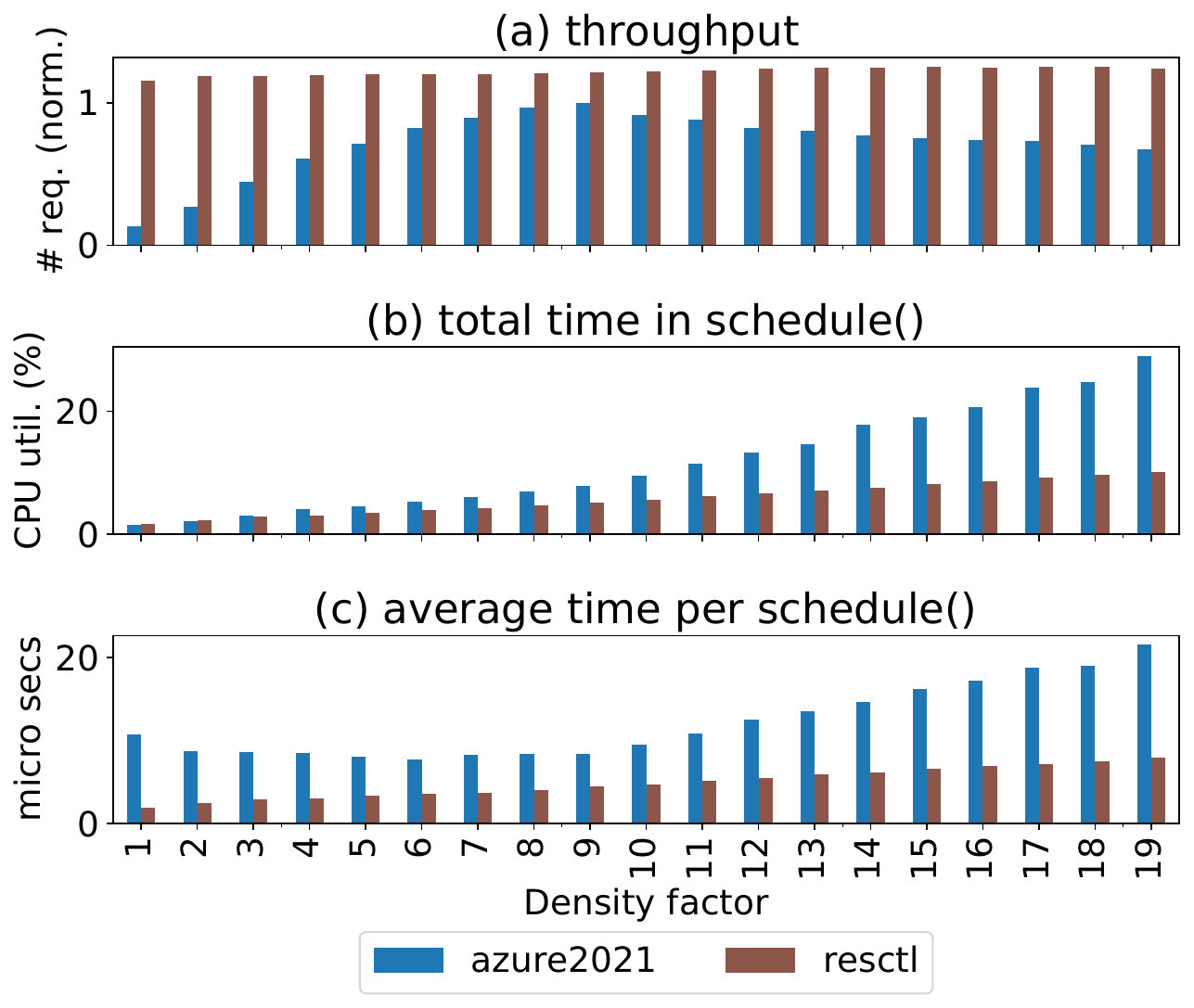}
  \caption{\label{fig:problem-overhead}Impact of increasing workload colocation
    on scheduling overhead.}
\end{figure}

We use Meta’s resource control framework~\cite{resctl} as the basis for
performing our evaluations. This uses cgroups configurations for targeted
workload simulators to replicate typical resource contention scenarios. The
primary workload simulator, \verb|rd-hashd|, emulates an online service that
continuously hashes data from disk, while also modeling the effects of
colocating opportunistic and background workloads. This allows it to saturate
all node resources: CPU, memory, storage. As we wish to focus on CPU overheads
only, we replace this with a simple CPU-bound Fibonacci computation.

We repurpose this framework to model the colocation of multiple
latency-sensitive functions by running multiple instances of \verb|rd-hashd| on
a single host.\footnote{We will release our framework as open source to support reproducibility.} Each instance represents a different function deployed as a
\verb|systemd| service with its own cgroup nested within a shared parent group
(\verb|faas.slice|), alongside the default user and system cgroups. For example,
on a 12\,core machine, the minimal configuration would run 12 instances
representing one function per core, a minimal cgroup setup compared to those
commonly found in production environments. Further experiments would then run
24, 36, $\ldots$ instances to scale up the load.

We drive the tool using the segmented traces described above, reflecting
realistic serverless request arrival patterns. The result is that we add a
trace-driven load scenario (\verb|azure2021|) alongside the default load mode
(\verb|resctl|) that self-tunes a steady-state concurrency level to meet a
target latency (e.g.,~100\,ms), representing a non-serverless workload in which
requests can be routed to other replicas.

Figure~\ref{fig:problem-overhead} presents the impact of increasing workload
colocation on both performance and scheduling overhead, comparing the default
load with the trace-driven load. Performance is evaluated in terms of
throughput, the number of requests completed within the standard latency target
of 1\,s~\cite{Swayam}.

In Figure~\ref{fig:problem-overhead}a, we increase workload colocation until maximum throughput is
achieved, and then further increase the number of functions beyond the system's
capacity to observe the effects of overload. This peak represents the point at
which the total demand from multiplexed workloads exhausts the available CPU
capacity. In our setup, the peak throughput of \verb|azure2021| is achieved
at a density factor of~9x (that is, the number of functions expressed as a multiple of hardware threads) giving 108 functions. Once workload colocation
exceeds this peak, throughput declines with fewer requests completing within the
latency target. At the highest level of workload colocation (i.e. density factor of~19), over 35\% fewer
requests meet the 1\,second latency target. This is in contrast to \verb|resctl|
where a higher peak throughput is maintained even under high workload colocation
due to the self-tuning concurrency which limits the number of contending
threads.

This significant throughput degradation for \verb|azure2021| can be explained by comparing the growth of
scheduling overhead (Figure~\ref{fig:problem-overhead}b), which includes the
total time spent in the core scheduler. At the point of peak throughput (i.e. density factor 9x),
scheduling overhead accounts for 5---7\% of CPU time, with no significant
difference between the trace-driven and default workloads.

However, the overhead for the \verb|azure2021| grows substantially beyond
the ~9x density---reaching up to 28\% at the highest level of colocation---compared to
less than 10\% for \verb|resctl|. This more dramatic increase in overhead
can be explained by the higher cost of an individual context switch for
\verb|azure2021| which can grow up to 20\,$\mu$seconds compared to less
than 10\,$\mu$seconds for \verb|resctl| (Figure~\ref{fig:problem-overhead}c).
This can be explained by the reduced level of queuing for \verb|resctl| where
new threads are spawned only after the completion of existing threads.

The fundamental observation is that {\bf the average cost of an individual context switch is
  higher under realistic arrivals and grows as the degree of colocation
  increases}.
Further analysis of \verb|ftrace| output shows that this is due to the additional
work the scheduler must perform when scheduling the next task to run from the
relevant runqueue, i.e.,~\verb|pick_next_task_fair| from \verb|fair.c|.\footnote{\url{https://elixir.bootlin.com/linux/v5.18.19/source/kernel/sched/fair.c\#L7278}} Each runqueue
is a red-black tree where the next task to be selected is always in the
left-most bottom-most node. {\bf Selecting the next \se via \verb|pick_next_entity|
  is thus efficient -- but the process of reinserting the preempted task and
  its descendants back into the red-black tree is not}.

Analysis of kernel function calls captured by ftrace shows that reinserting preempted tasks involves multiple
calls to \verb|put_prev_entity|,  increasing the cost of task scheduling
by several times (dozens of microseconds). The overhead becomes increasingly significant as context switching occurs between tasks that are not siblings within the same cgroup, and as the context switch rate increases.

\subsection{Overhead in cluster-mode evaluation}\label{sec:motivation:cluster}

Finally, we examine the core scheduler's overhead in a more realistic cluster scenario
using the \emph{cluster node} mode, running  real function code invoked via the control
plane of the serverless framework---in this case, Knative (see Figure \ref{fig:knative-cgroups}). We drive this scenario using a trace-driven workload generator that triggers function code executing multiple instances of a PyTorch image classification model. This model is wrapped using the BentoML serving framework to expose an HTTP API, with all instances sharing CPU resources on a single server managed by Knative. This setup performs I/O operations such as receiving and resizing images, and it results in a multi-threaded function structure with two thread pools for queuing and batching requests, plus a separate thread pool for model serving. This approach is widely reported to maximize the opportunity for vertical scaling by increasing function concurrency and reducing the effects of cold starts~\cite{Golgi, FaaSNet, GoogleCloudRunConcurrency, HuaweiTraces}.

The workload generator invokes the PyTorch function via Knative's API ingress, which is
proxied by Knative's sidecar container. The workload generator thus behaves like
a pseudo open-loop system, as requests may be queued via Knative's Istio ingress
when the node becomes overloaded. Overhead is measured given 100 colocated functions which
reflects the maximum that can be colocated on a single Kubernetes node, based on
the default limit of 110 pods/node, with 10
pods reserved for essential system services~\cite{k8s-node-limit}.

Under such load conditions, CPU scheduling overhead rises to approximately 20\%
of total CPU time, with an average context switch of 48\,$\mu$seconds,
significantly higher than in the standalone host case (i.e. 10---20\,$\mu$seconds). This increase is
primarily attributed to two factors: \one~more frequent queuing due to the
handling of HTTP requests and responses between the sidecar proxy and the ML
serving framework, and \two~longer execution times that result in CPU saturation
with fewer concurrently active functions.

\textit{\textbf{Conclusion.} Higher workload colocation makes serverless workloads vulnerable to performance degradation due primarily to the rising cost and frequency of context switching. As a result, the total spent on context switching grows disproportionately relative to the amount of useful work being performed.}

\section{Latency-Aware Group Scheduling (LAGS)}
\label{sec:cfs-llf}

This section introduces CFS-Latency-Aware Group Scheduling (\cfsllf), a novel approach to Linux CPU scheduling that improves group scheduling in high-density serverless deployments. The design builds upon Linux group scheduling~\cite{cfs-group}, which itself is implemented on top of CFS~\cite{CFS}.

\cfsllf employs the cgroup interface to schedule at the granularity of multi-threaded serverless functions. For instance, it may be applied to schedule Knative functions (see Figure \ref{fig:knative-cgroups}), where Kubernetes pods encapsulate a user container, its sidecar proxy, and any additional containers. These all contribute to a function's total execution time.

\begin{figure}
  \centering
  \includegraphics[width=\linewidth]{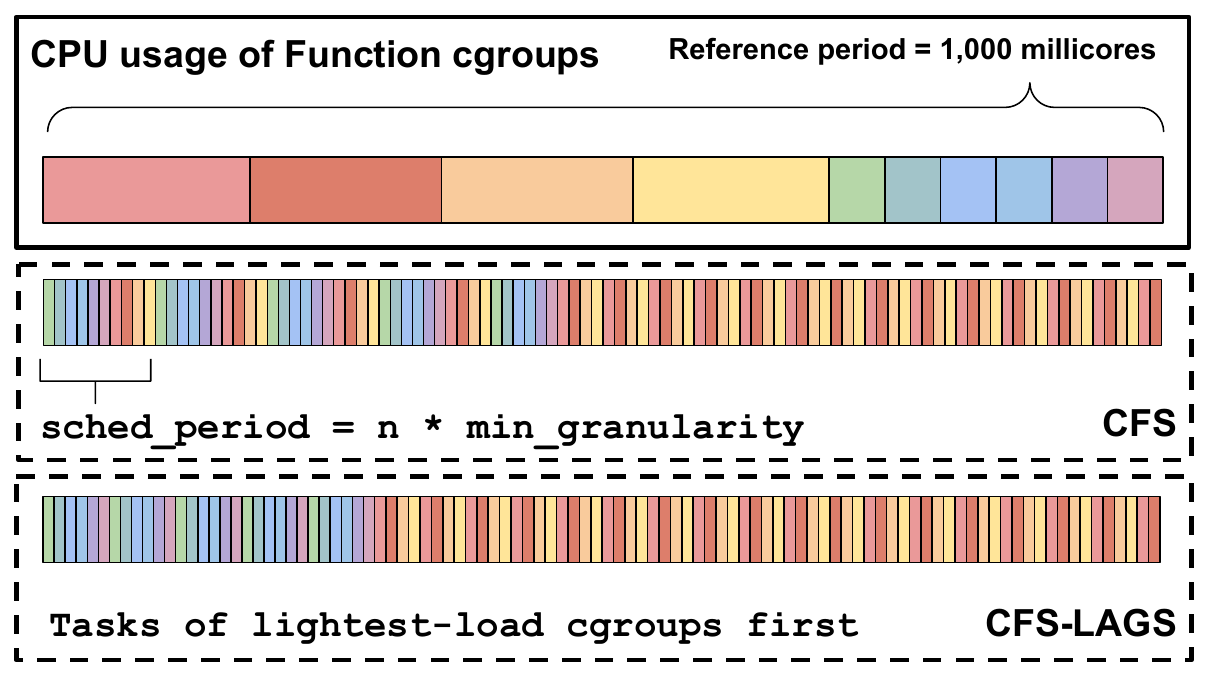}
  \caption{\label{fig:cfs-llf-conceptual} \cfsllf's cgroup-aware scheduling in comparison with the default group scheduling in CFS.}
\end{figure}

The fundamental insight of \cfsllf is to maximise task completion within function cgroups over a reference period that extends beyond the scheduler’s short scheduling interval. Figure~\ref{fig:cfs-llf-conceptual} presents a high-level conceptual illustration of cgroup-aware task completion for a simple one-millicore configuration with a single level of cgroup nesting. Under the default CFS policy, fairness is enforced by scheduling threads from all cgroups equally within each scheduling period. By contrast, \cfsllf takes account of the load contributed by each cgroup and prioritises tasks from the lightest-loaded groups. This approach not only reduces median turnaround time but also allows such cgroups to exit the system earlier, shortening run queues and mitigating  associated overheads.

\textbf{\cfsllf goals and properties.}
The goal of \cfsllf is to utilise task completion to reduce scheduling overhead in environments characterised by high colocation of cgroup-managed, multi-threaded workloads~\s{sec:motivation}. The remainder of this section sets out how this goal is realised as follows:
\begin{itemize}
\item We first motivate the underlying scheduling principles using a static approach, demonstrating how latency-aware scheduling can mitigate contention in high-density scenarios~\s{sec:cfs-llf-static}.
\item We then describe how \cfsllf achieves latency-awareness with cgroup-aware task completion using the \emph{Load Credit} metric, which tracks recent CPU usage for all threads within a function~\s{sec:cfs-llf-loadcredit}.
\item We conclude with the structural changes required to realise \cfsllf, including per-cgroup policy adjustment and policy-aware task-to-core placement~\s{sec:cfs-llf-implementation}.
\end{itemize}

\subsection{Motivating Lightest Load First approach}
\label{sec:cfs-llf-static}

\begin{figure}
  \centering
  \includegraphics[width=1\linewidth]{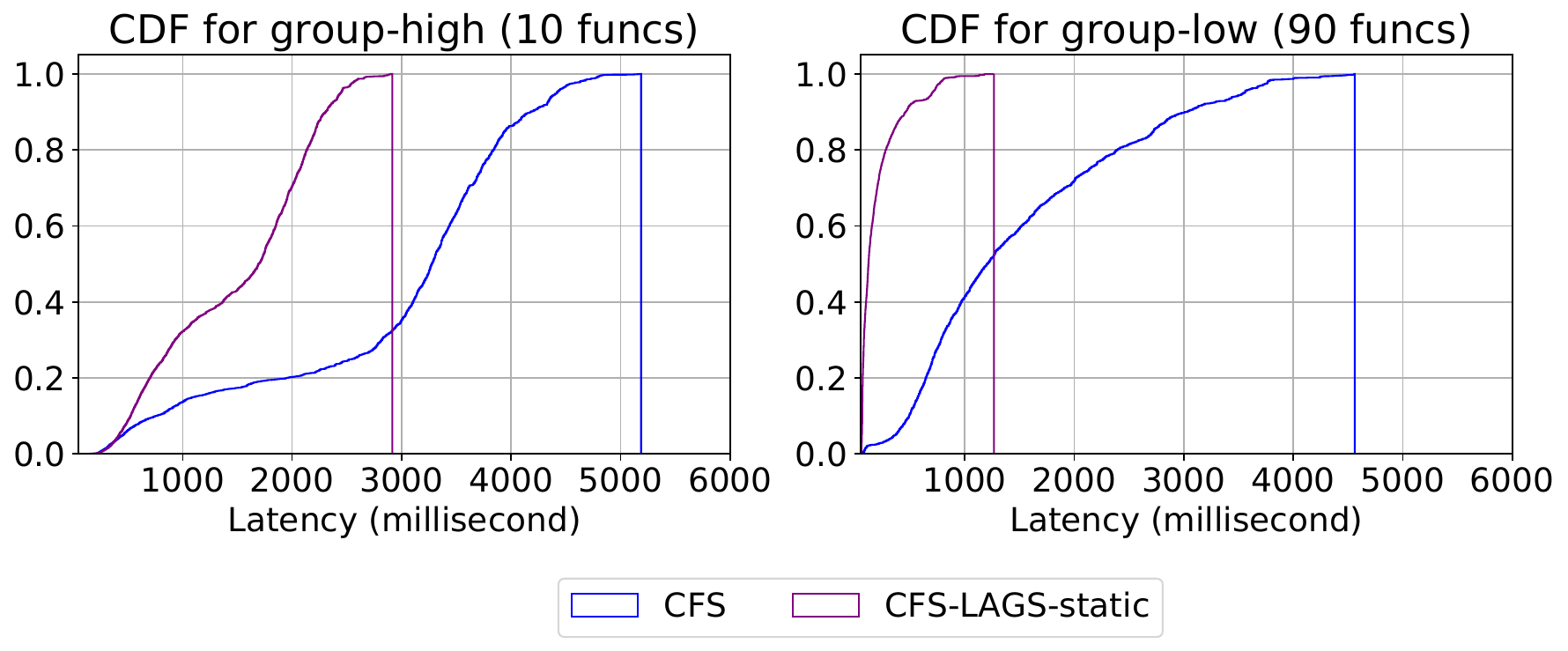}
  \caption{\label{fig:worst-contention-cdf}Latency CDFs under the highest
    colocation of 100\,pods/node for the microbenchmark cluster
    mode~\s{sec:motivation}.}
\end{figure}

This section motivates our scheduling approach based on giving priority to functions with the lightest load. We demonstrate that with  \cfsllf-static which uses the round-robin Linux real-time scheduling policy (\verb|SCHED_RR|) to prioritise a subset of colocated functions, thereby statically realising latency-awareness and measuring its potential to reduce contention under large colocation. \cfsllf-static relies on offline knowledge of future demand—an infeasible assumption in practice—to validate our core insight, implemented with the \verb|chrt| tool, and serves as an evaluation baseline to inform \cfsllf tuning \s{sec:cfs-llf-loadcredit} and implementation \s{sec:cfs-llf-implementation}.

\cfsllf-static statically prioritises
a number of functions belonging to the lowest demand band from the trace-driven colocation benchmark in~\S\ref{sec:motivation} (Figure~\ref{fig:skewed-demand}).
The \verb|SCHED_RR| policy allows for a maximum quantum of 100 ms per function, while the rest remain under the default CFS policy.
Tasks under \verb|SCHED_RR| can immediately preempt other CFS tasks and by
default can take up to 95\% of the one-second scheduling period. This results in
a hybrid scheduler that services low load tasks (\emph{group-low}) quickly,
allocating any remaining time to high load tasks (\emph{group-high}). The ideal
number of functions for group-low is obtained experimentally by adding new functions to
the group as long as there is no measurable impact on the latency target of
1\,second.

Figure \ref{fig:worst-contention-cdf} illustrates the CDFs of request latencies
for the two groups of functions under the two scheduling approaches, CFS and
\cfsllf-static. In the case of group-low, \cfsllf-static significantly reduces
the tail of the latency distribution due to the high preemption priority of this
group under \verb|SCHED_RR|. Perhaps counter-intuitively, this also leads to a
significant improvement for group-high over the baseline CFS even though
this group remains scheduled under the same CFS policy in both setups.

Statistics collected using \emph{schedstats}
tool~\cite{schedstats} show that the \cfsllf-static leads to a reduction
of over 75\% in the total time tasks spend waiting in CPU run queues, which
accounts for the overall improvement in latency. The CPU is also left idle almost twice as often, reflecting improved
efficiency due to freed-up compute cycles.

\subsection{Realising \cfsllf with the Load Credit metric}
\label{sec:cfs-llf-loadcredit}

\begin{figure}
  \centering
  \includegraphics[width=1\linewidth]{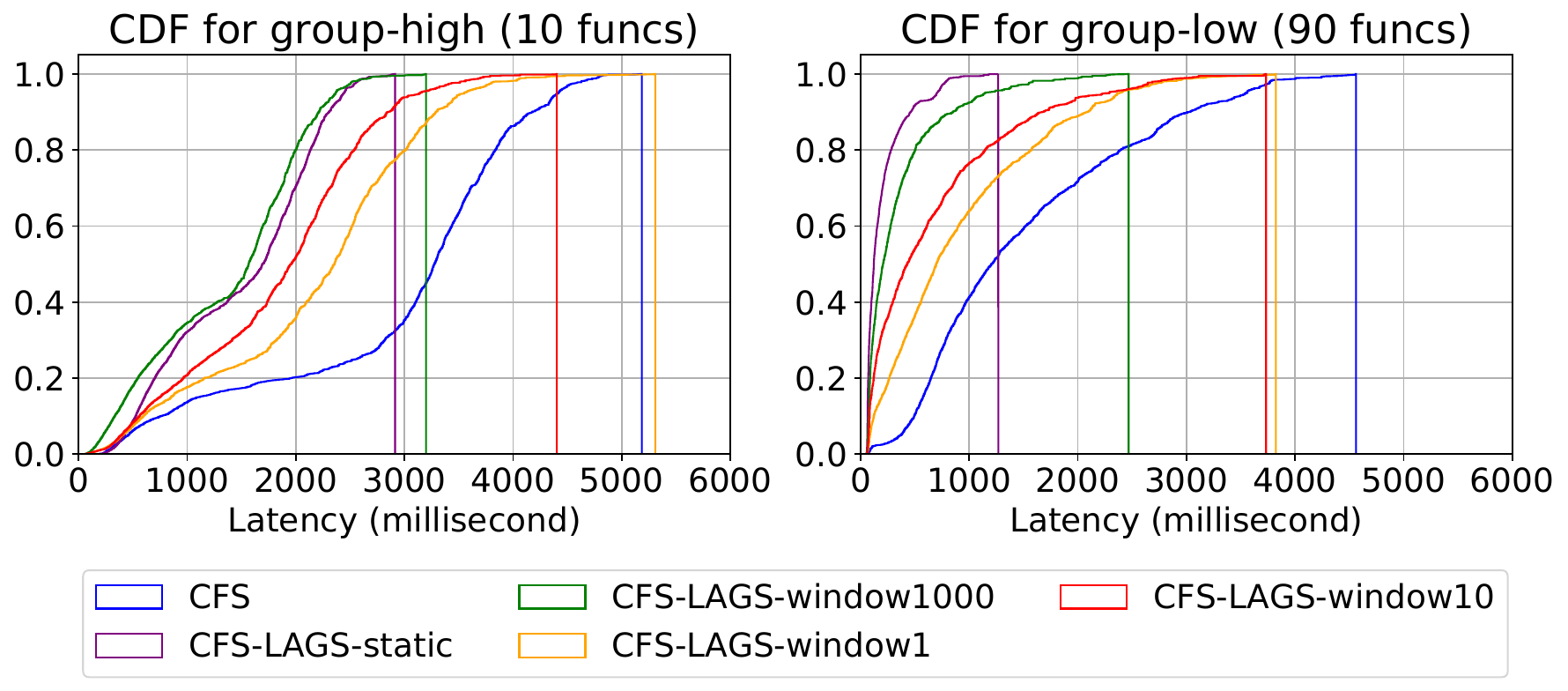}
  \caption{\label{fig:sensitivity-analysis} Impact of the Load Credit
    window size on approximating the static \cfsllf policy.}
\end{figure}

We now introduce the \emph{Load Credit} metric, which is used by \cfsllf to realise latency-awareness online for colocated functions. This enables a practical realisation of the benefits of \cfsllf-static without requiring knowledge of the future. This metric captures the CPU usage of a function over an extended period of time (on the order of a few seconds). \cfsllf operates by prioritising cgroups with the \emph{lowest Load Credit}, thereby relaxing strict CFS fairness by allowing threads of a function to continue running as long no other cgroup has a lower Load Credit. In addition to improving \emph{invocation latency}, this approach also provides \emph{resilience} under overload: CPU cores continue performing useful work even in the presence of contention, by prioritising tasks from cgroups that contribute the least to overall contention, rather than enforcing strict fairness among all competing tasks.

The Load Credit metric is tracked at the granularity of cgroups corresponding to serverless functions and is used to manage the scheduling of the associated group scheduling entities (\verb|gse|). For example, in our Knative setup (Figure~\ref{fig:knative-cgroups}), \cfsllf uses the Load Credit for \verb|kubepods-{1..100}| to manage scheduling for the \verb|cfs_rq| run queues corresponding to \verb|kubepods-burstable|. Scheduling for \verb|cfs_rq| run queues of  other cgroups continues under the default CFS policy.

Using the Load Credit metric makes \cfsllf comparable to the Least Attained
Service~(LAS) policy~\cite{Kairos}, a variant of SRPT that does not depend on
runtime estimates for the remaining service execution time. LAS schedules the
\emph{youngest} task, that with the least attained service, for
execution. If multiple tasks have the same attained service, they share
processing time equally, similar to a Processor Sharing (PS) policy. LAS assumes
that the amount of service a task has consumed is a good indicator of its
remaining service demand, making it particularly effective with heavy-tailed
service demand distributions where larger tasks typically take more time to
complete.

\textbf{Tuning the Load Credit size}
The Load Credit metric is realised in CFS group scheduling by building upon the
PELT load tracking mechanism~\cite{PELT}, used for cross-core load balancing.
The key idea is to reuse the Linux task-level load metric aggregated at the cgroup level, which
corresponds to function sandboxes across all cores.
In vanilla CFS, the PELT
metric is readily accessible, as each cgroup maps to a task group data
structure, allowing access to the load metric for the corresponding cgroup
through \verb|tg->load_avg|. Load Credit is calculated as an exponentially
moving average of PELT over a larger time
window and stored in a new attribute (\verb|tg->load_avg_ema|). With this new metric tracked over an appropriate time window, it becomes
possible to identify and prioritise  those functions contributing least to the system
load within CFS at
runtime.

The Load Credit size value is critical for improving task completion rates and addresses the overhead issue~\s{sec:cfs-llf-evaluation}. Prioritising functions over a larger window allows them to execute for longer, capturing more substantial variations in request bursts based on the function’s recent activity. The Load Credit size determines how quickly a function can continue to execute before its Load Credit increases, and other tasks are then prioritised.
Figure~\ref{fig:sensitivity-analysis} illustrates the
impact of various configurations of this parameter on latency under the cluster
node 100 function colocation scenario discussed in~\S\ref{sec:cfs-llf-static}. A
window size of 1,000 scheduler ticks (\verb|CFS-LAGS-window1000|) yields the
best results, roughly equivalent to 4\,seconds given the default
kernel configurations and a kernel timer frequency of
\texttt{CONFIG\_HZ=250}.
Within this window, \cfsllf dynamically prioritises
tasks belonging to least-loaded functions as though they were
allocated using \verb|SCHED_RR|, without risking them taking over the
system.

\subsection{Implementation}
\label{sec:cfs-llf-implementation}

The implementation of \cfsllf requires non-trivial modifications to CFS’s highly concurrent codebase due to its per-cgroup, per-core run queu structure~\s{sec:cfs-background}. \cfsllf is realised as a Linux kernel patch\footnote{\url{https://codefile.io/f/73L5WeDAlk}} developed against version 5.18, the most recent stable kernel available at the outset of this work. The patch is primarily based on the  \textbf{default CFS scheduler class} in \verb|kernel/sched/fair.c|, which is one of the oldest and most complex components of the kernel having over 10,000 lines of code.  The patch modifies relatively stable scheduler code paths, making it straightforward to port to later kernel versions. In total, \cfsllf adds just under 300 lines of code in \verb|kernel/sched/fair.c| plus introduction of new attributes into existing CFS data structures.

\textbf{Fine-grained per-cgroup policy adjustment.}
In \cfsllf, the Load Credit mechanism governs the scheduling policy for group scheduling entities (\verb|gse|) corresponding specifically to function cgroups, whilst preserving the existing CFS policies for all other run queues. Such fine-grained control is not supported by existing frameworks for modifying CFS~\cite{sched-ext, Enoki, Plugsched}, which apply the same scheduling policy across all cgroup run queues. The standard CFS code is relatively straightforward as it simply compares the \verb|vruntime| of two scheduling entities to decide which to preempt. However, incorporating Load Credit into this comparison proved non-trivial, as the context in which it occurs is highly complex. CFS data structures are traversed and initialised through multiple code paths, in varying orders—bottom-up during task wake-ups and top-down when scheduling the next task.

\textbf{Policy-aware task-to-core placement.} In \cfsllf, task completion is ensured across all cores—an objective absent from the original CFS design, which focused on load balancing and work-conservation~\cite{WastedCores,Ipanema}.
In earlier versions of \cfsllf, considering Load Credit only at the level of per-CPU run queues was insufficient, as the effects of load balancing were also critical.
To address this, we incorporate the Load Credit metric into the load-balancing code path, effectively acting as a greedy heuristic to identify a suitable, preferably idle, target CPU core. We further extend these heuristics to ensure that a newly woken task is placed on the first available CPU core that is either idle or currently executing a task with a lower Load Credit.
This approach is inspired by the \verb|SCHED_IDLE| fix patch~\cite{cfs-schedidle}, which enables latency-sensitive tasks to pre-empt low-priority tasks under \verb|SCHED_IDLE| by more aggressively assigning them to cores running such low-priority workloads.

Appendix~\S\ref{sec:patch} details the user-space interface to \cfsllf and the changes required to incorporate the Load Credit metric into per- and cross-core scheduler code paths.

\section{Evaluation}
\label{sec:cfs-llf-evaluation}

We earlier demonstrated that \cfsllf significantly improves the performance of a single node running colocated serverless functions~\s{sec:cfs-llf}. Our evaluation focuses first on end-to-end cluster-level performance gains as the most important overall measure~\s{sec:cluster-case-study}. As the scheduler is a core kernel component, it is important that any modifications to better support a particular workload---in this case, serverless---are not to the detriment of other workloads, so we also examine how \cfsllf performs under a range of alternative extreme workload scenarios beyond serverless~\s{sec:host-evaluation}.

\textbf{Workload.} Our evaluation leverages our benchmarking framework~\s{sec:motivation}, with the cluster case study~\s{sec:cluster-case-study} extending the cluster mode~\s{sec:motivation:cluster} using realistic workload conditions based on the advanced cgroup setup for Knative (see Figure~\ref{fig:knative-cgroups}) and real function code. This allows us to capture the benefits of \cfsllf when deployed in a production-like cluster environment. For regression analysis~\s{sec:host-evaluation}, we extend the use of the
stand-alone host mode~\s{sec:motivation:host}, applying synthetic
load scenarios with a basic cgroup setup
(i.e.,~\verb|faas.slice/| \verb|func-{0..n}.service|) to evaluate the benefits of \cfsllf in  controlled conditions with minimal cgroup  configuration.

\textbf{Baselines}. We believe that as group schedulers only CFS and EEVDF are directly comparable to \cfsllf. We retain CFS as our primary baseline, as it shares an identical internal structure with \cfsllf, enabling a more precise attribution of performance improvements to our modifications. We discuss the other work-in-progress kernel patches that attempt to address group scheduling inefficiencies of which we are aware~\cite{cfs-flattened, Hierarchical-CBS} later~\s{sec:related-works}.
Schedulers designed specifically for serverless workloads (e.g.,~SFS~\cite{SFS}, ALPS~\cite{ALPS}) operate on individual invocations via ephemeral processes and do not extend to cgroup-managed multi-threaded workloads as our discussion of cgroup-aware task completion illustrates~\s{sec:cgroup-task-completion}.

\textbf{Evaluation setup}. All single host experiments are run on a dedicated server with an Intel Xeon
CPU E5-2430L having six physical cores with two hardware threads per core and 64\,GB of
memory. This single socket setup is essential to enable a focused development and
evaluation of \cfsllf independently from
cross-socket load balancing mechanisms. Such mechanisms involve orthogonal concerns such as
energy management~\cite{Nest} and hardware heterogeneity which is left for
future work.

We then reproduce all experiments from~\s{sec:host-evaluation} on a multi-socket server to validate the portability of \cfsllf to high-end systems. This server is based on an AMD EPYC 7453 processor with two 28-core sockets, offering a total of 112 hardware threads across 8 NUMA nodes. We observe that with static cross-node load balancing (pinning an equal number of functions to each NUMA node using \verb|cpusets|), the behaviour of \cfsllf remains comparable to the results reported in~\S\ref{sec:host-evaluation}. However, when tasks are load balanced across multiple NUMA nodes, the observed improvements are present but with only 50\% of the reported improvements. This suggests that further integration of the Load Credit metric into cross-node  scheduling is feasible.

All cluster experiments are conducted on a Kubernetes v1.23 cluster, using identical nodes to those in the the single-socket host experiments, with \verb|containerd| v1.4.12 as the container runtime and the Knative control plane. Management and control plane functions are hosted on dedicated nodes
rather than the worker nodes used for measurement.

\subsection{End-to-end cluster performance}\label{sec:cluster-case-study}

We examine a cluster case study to demonstrate how \cfsllf
facilitates higher-density serverless deployments by mitigating the overhead
associated with host-level scheduling. Using a downscaled version of the Azure
Functions Trace 2019~\cite{Shahrad}, we evaluate performance across a cluster of
15 dedicated servers (180 hardware threads in total) running the PyTorch image
classification function of the cluster mode microbenchmark~\s{sec:motivation}.
The Azure workload is proportionally scaled to fit within this cluster size,
preserving the original temporal distribution which results with a sample of
around 800 containers. Baseline performance is established by profiling the
functions in an oversized cluster to eliminate contention, revealing that 14
nodes are required to meet peak demands through static resource reservation.

\begin{figure}
  \centering
  \includegraphics[width=1\linewidth]{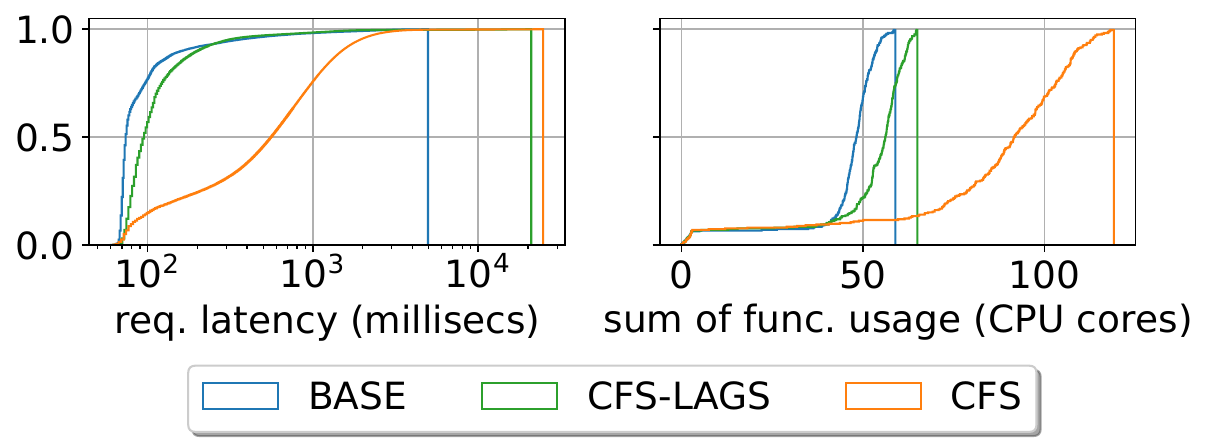}
  \caption{\label{fig:colocation-impact}Cluster-wide CDFs of latency (left) and
    CPU utilisation (right) for the case study~\s{sec:cluster-case-study}.}
\end{figure}

Guided by this profiling data, we continuously consolidate the workload onto fewer
\cfsllf-enabled nodes that maintain the same performance which
results with a cluster setup with just 10 nodes---a 28\% reduction in server
capacity---without compromising performance
(Figure~\ref{fig:colocation-impact}~(left)). In contrast, using the default
Linux CFS scheduler under identical conditions results in up to a
6$\times$ increase in median and tail latencies due to the significant amount of CPU time wasted context switching in
the kernel scheduler.

The host scheduling overhead can be measured at the cluster level in the form of
clear disparity between effective and perceived CPU utilisation
(Figure~\ref{fig:colocation-impact}~(right)). Profiling data indicates that peak
usage across all functions in an over-provisioned cluster is around 60 cores.
With \cfsllf, the gap between perceived and actual usage is minimal---around
+10\% ($\sim$65 cores)---compared to a substantial +100\% overhead under CFS
($\sim$120 cores). This gives the misleading impression of full
utilisation while significant CPU cycles are in fact being wasted context
switching.

Production case studies report that CFS requires the target utilisation to be no
more than 45\%~\cite{AlibabaOC}. Our experiments, show that \cfsllf allows average CPU utilisation to rise
to 55\%, enabling operators to safely increase server density while maintaining similar performance---something that is not achievable
under the standard CFS scheduler.

\subsection{Host efficiency}
\label{sec:host-evaluation}

We evaluate \cfsllf using workloads that extend the stand-alone host
microbenchmark~\s{sec:motivation:host},
highlighting key scenarios in the workload space that are possible in a high-density deployment.
We first systematically vary function colocation density
and invocation arrival patterns~\s{sec:host-evaluation:arrivals},
illustrating the impact of context switching overhead~\s{sec:host-evaluation:overhead}.
We then vary workload parallelism and execution times to demonstrate the
advanced task completion capabilities of \cfsllf~\s{sec:cgroup-task-completion} in comparison to other baselines.

\subsubsection{Performance across varying load and arrival patterns}\label{sec:host-evaluation:arrivals}

We use constant load (\verb|resctl|) and realistic arrivals (\verb|azure2021|) introduced earlier in \s{sec:motivation}.
Constant load (resctl) represents the best-case scenario with minimal scheduling overhead, reflecting an ideal steady state of ``serverful'' workloads comprising long-running functions without significant gaps between invocations. Realistic arrivals (azure2021), on the other hand, capture interactive, co-located workloads while assuming overlapping peak loads similar to Azure colocation scenarios.
In addition, we also introduce random arrivals (\verb|random|) which represents the worst-case request
arrival pattern for host scheduling with a large number of small functions.
Each function has between 0---5 invocations per second, uniformly drawn, with peak
aggregate demand matching that of \verb|azure2021|.

\begin{figure}
  \centering
  \begin{subfigure}{0.49\textwidth}
    \includegraphics[width=\textwidth]{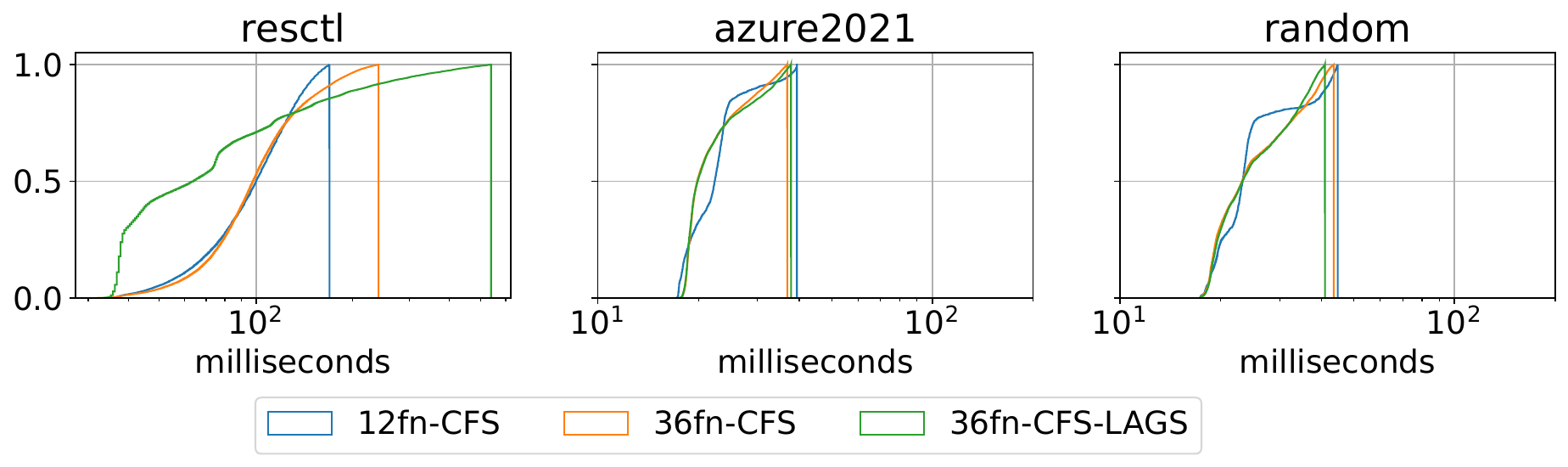}
    \caption{\label{fig:solution-cdfs-low}Density factor 3 (low load).}
  \end{subfigure}
  \hfill
  \begin{subfigure}{0.49\textwidth}
    \includegraphics[width=\textwidth]{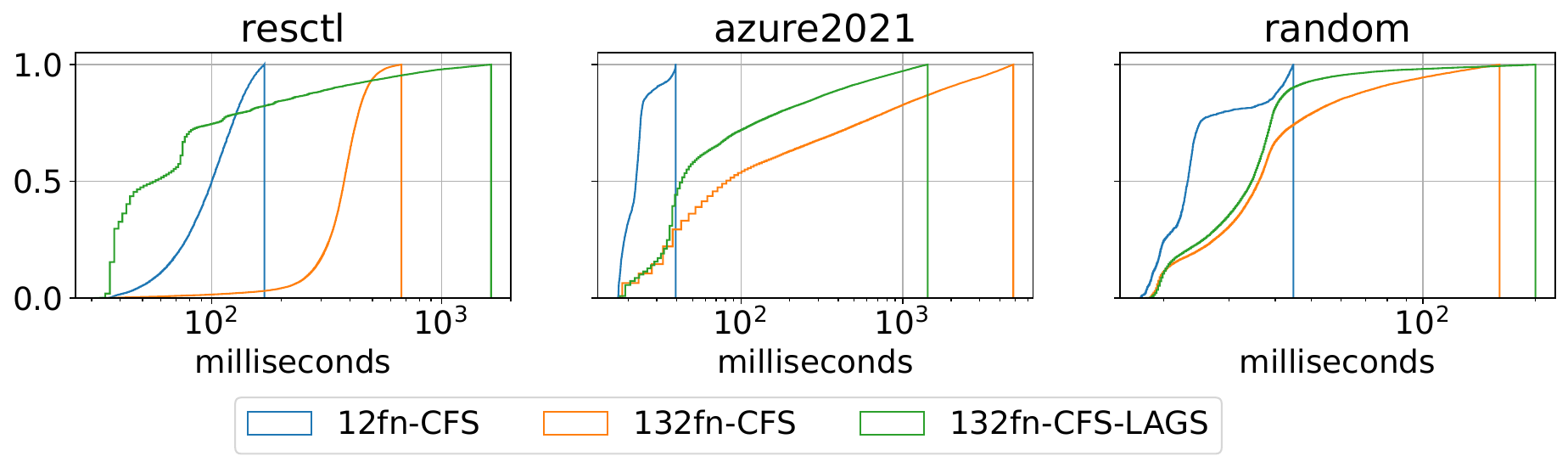}
    \caption{\label{fig:solution-cdfs-high}Density factor 11 (high utilisation).}
  \end{subfigure}
  \hfill
  \begin{subfigure}{0.49\textwidth}
    \includegraphics[width=\textwidth]{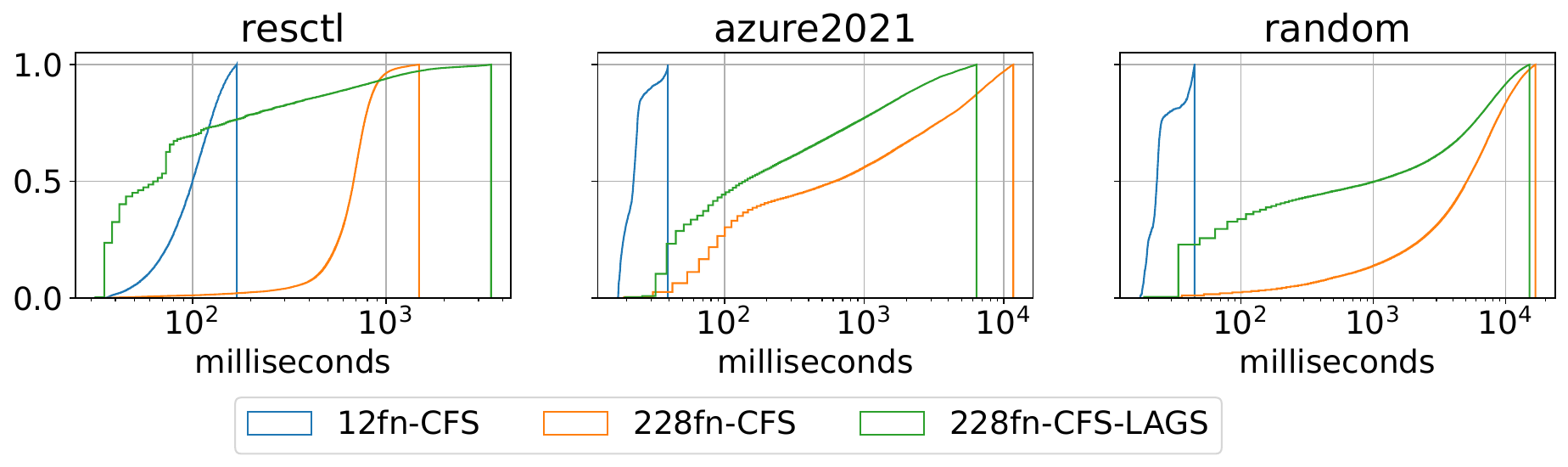}
    \caption{\label{fig:solution-cdfs-overload}Density factor 19 (overload).}
  \end{subfigure}
  \caption{\label{fig:solution-cdfs}Latency CDFs given selected colocation
    scenarios. Note the log-scale $x$-axis.}
  \vspace*{-3mm}
\end{figure}

We run the three workloads under
increasing colocation levels.
Figure~\ref{fig:solution-cdfs} shows latency CDFs for the workloads at
selected density levels corresponding to low utilisation (36 functions, i.e.,~3 per core), high
utilisation (132 functions, i.e.,~11 per core), and significant overload (228 functions, i.e.,~19 per core). The
common baseline across all these is the latency distribution of CFS when the
number of functions (12) equals the number of hardware threads (12).

Figure~\ref{fig:solution-cdfs-low} shows latency distributions under low load,
and we see that \cfsllf and CFS are comparable for this workload. However, we
can also see via \verb|resctl| in
Figures~\ref{fig:solution-cdfs-low},~\ref{fig:solution-cdfs-high}
and~\ref{fig:solution-cdfs-overload} that \cfsllf introduces noticeable tail
latency for constant load across all load scenarios.

This demonstrates a fundamental trade-off that is inherent to \cfsllf. Under the
steady load of \verb|resctl|, \cfsllf exhibits dynamics similar to SRTF: median
latency is significantly reduced at the expense of increased tail latency.
However, as the CDFs show, the majority of requests below the $95^{th}$ percentile
improve in comparison to the default CFS policy. In contrast, the strict
fairness in CFS enforced at very short time scales reduces tail latency at the
expense of impacting median latency for all requests.

Tail
latency is an issue under constant load but more realistic bursty workloads, e.g.,~\verb|azure2021|,  have gaps between requests that mitigate this tail latency
issue.
\cfsllf significantly reduces this tail under high load (\verb|azure2021| in Figures~\ref{fig:solution-cdfs-high}
and~\ref{fig:solution-cdfs-overload}). These results are further discussed in \S\ref{sec:host-evaluation:overhead}.
For the \verb|random| workload, tail latency under \cfsllf remains comparable to CFS across all scenarios.

Finally, Figure~\ref{fig:solution-cdfs-overload} shows results for the worst case
scenario where the host is overloaded with 228 functions. This case exhibits a
significant performance degradation for CFS, while \cfsllf significantly mitigates
the increased median latency across all workloads. Such an overload case should
only arise as a result of low quality placement decisions by the cluster
scheduler, ultimately requiring migration or re-scheduling some functions to
other nodes to resolve. However, it does help illustrate how \cfsllf reduces the impact of
overhead, ensuring the CPU continues to do useful work for longer instead of
wasting cycles context switching.

\begin{figure}
  \centering
  \includegraphics[width=\linewidth]{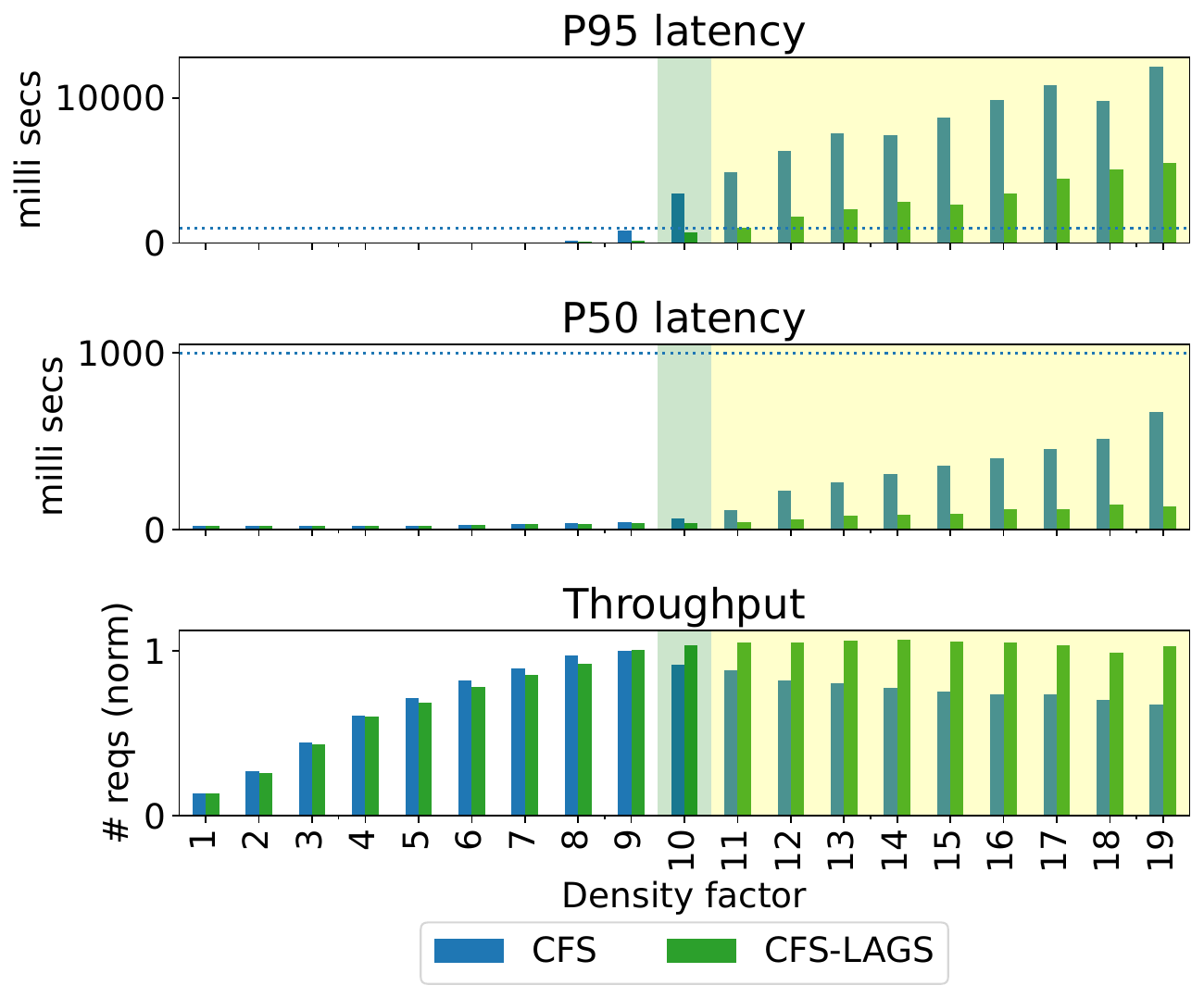}
  \caption{\label{fig:colocation-performance}Performance as function colocation increases under realistic arrivals (\texttt{azure2021}). Pale green highlight indicates the ideal density factor to avoid exceeding the latency target of 1\,s, indicated by the horizontal dotted line.}
\end{figure}

\subsubsection{Scheduling overhead and contention resilience}\label{sec:host-evaluation:overhead}
Reducing tail latency under realistic workloads helps increase function colocation for a given target latency.
Figure~\ref{fig:colocation-performance} compares the performance of CFS and
\cfsllf for \verb|azure2021| as colocation increases.

Performance is described in terms of median and tail latencies as well as the
throughput in terms of the number of requests which are executed within the
latency target of 1\,second. While increasing workload colocation does not cause
the median latency to exceed this target, the tail latency does exceed it under
CFS beyond a density factor of~8$\times$ (96 functions). This represents the ideal
number of functions to be colocated for CFS. In contrast, under \cfsllf it is
possible to accommodate at least another 12 functions without exceeding the same
latency target. This improvement can be significant in practice given the larger
overhead of real interactive functions as we demonstrated~\s{sec:cluster-case-study}.

\begin{figure}
  \centering
  \includegraphics[width=\linewidth]{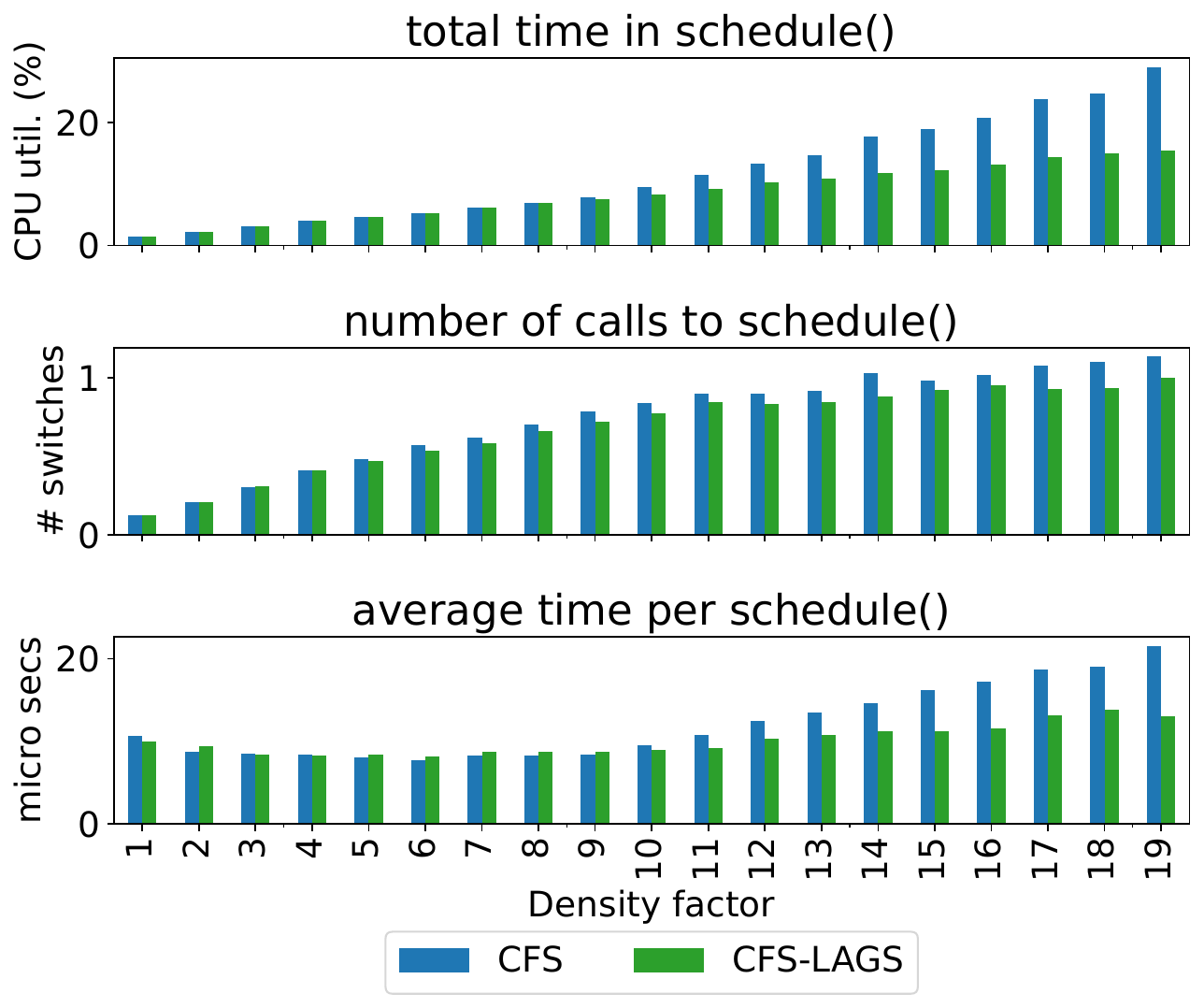}
  \caption{\label{fig:colocation-overhead}Scheduling overhead with increasing
    colocated functions under realistic arrivals (\texttt{azure2021}).}
\end{figure}

Furthermore, even when this ideal threshold for workload colocation is exceeded,
\cfsllf mitigates the impact on median and tail latency, reducing degradation in
system throughput from 35\% to less than 10\%. By thus reducing the severity of
impact of server overload, \cfsllf makes it feasible to use more aggressive
workload colocation policies, by decreasing scheduling overhead, as shown in
Figure~\ref{fig:colocation-overhead}. While \cfsllf reduces the frequency of
context switching by less than around 13\% the improvement can be mainly
attributed to reducing the average cost of a single context switch which drops
from 21\,$\mu$seconds to around 13\,$\mu$seconds.

\subsubsection{Cgroup-aware task completion in \cfsllf}\label{sec:cgroup-task-completion}

\begin{figure}
  \centering
  \begin{subfigure}{0.49\textwidth}
    \includegraphics[width=\textwidth]{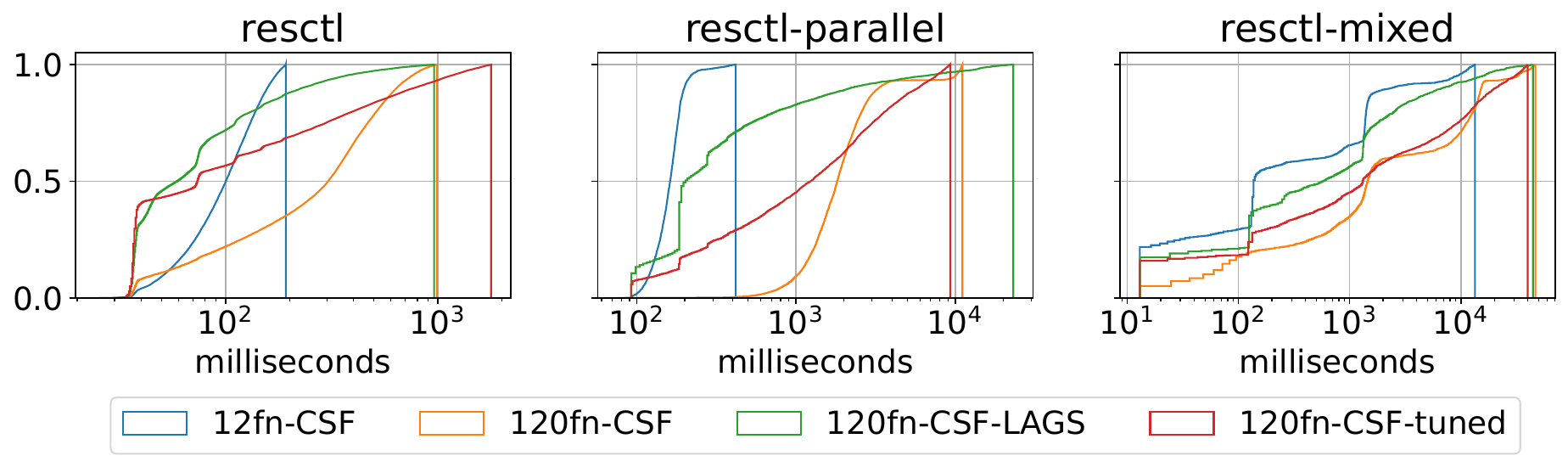}
    \caption{\label{fig:baselines-cfstuned}Linux CFS, with and without tuning}
  \end{subfigure}
  \hfill
  \begin{subfigure}{0.49\textwidth}
    \includegraphics[width=\textwidth]{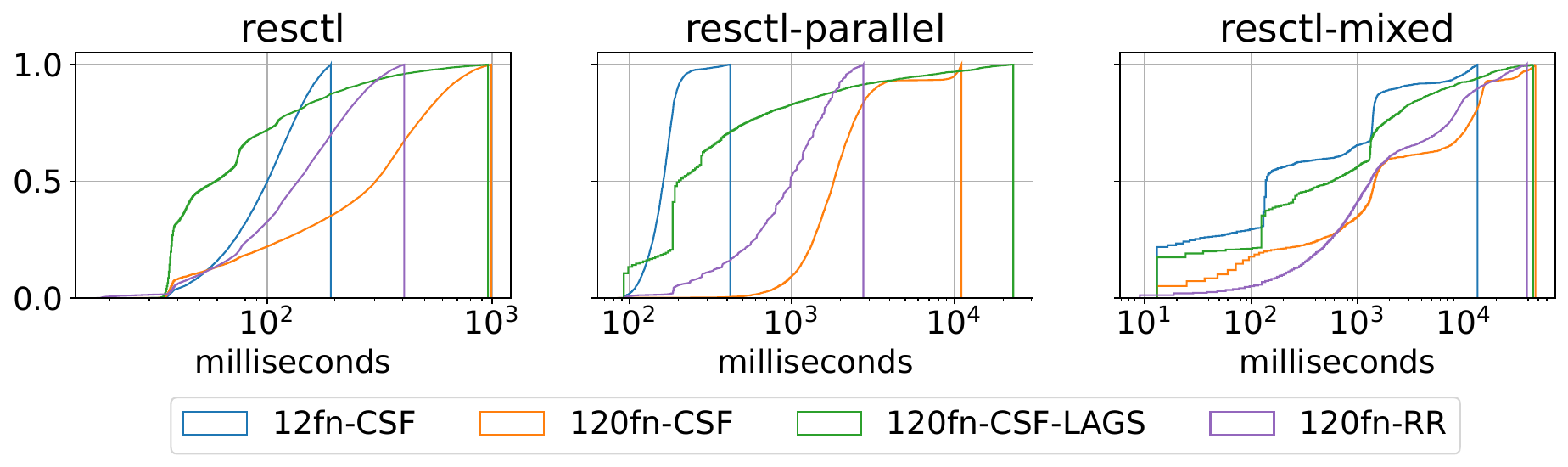}
    \caption{\label{fig:baselines-rr}Linux RR}
  \end{subfigure}
  \hfill
  \begin{subfigure}{0.49\textwidth}
    \includegraphics[width=\textwidth]{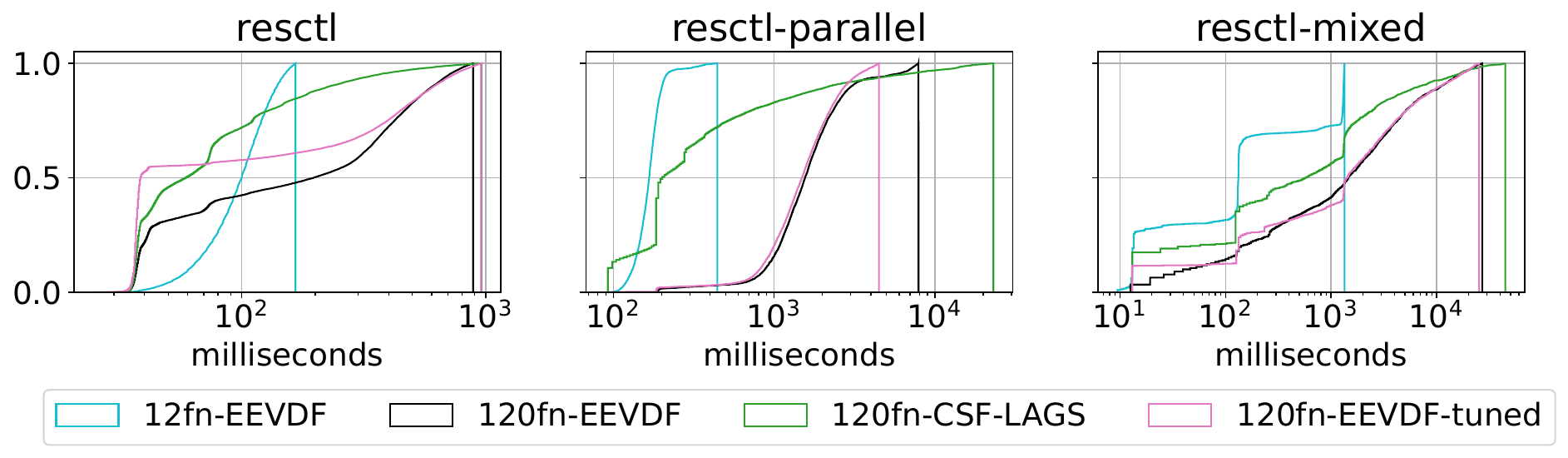}
    \caption{\label{fig:baselines-eevdf}Linux EEVDF, with and without tuning}
  \end{subfigure}
  \caption{\label{fig:baseline-cdfs}Comparison of latency CDFs of \cfsllf to evaluation baselines in \s{sec:cgroup-task-completion}. Note the log-scale $x$-axis.}
\end{figure}

Finally, we demonstrate how \cfsllf enhances task completion via use of cgroups, comparing with the other scheduling options that can be tuned to increase task completion by increasing the default task time slice.
These include the EEVDF~\cite{EEVDF-linux} scheduler from the most recent stable kernel, v6.12, and the round-robin soft real-time policy (Linux RR) which guarantees each function a default time slice of 100\,ms. We tune both CFS and EEVDF to match that time slice.

Figure~\ref{fig:baseline-cdfs} shows the results of evaluating the baselines against the default
\verb|resctl| workload, in addition to two additional workloads, \verb|resctl-parallel| and \verb|resctl-mix| that
highlight advanced task completion cases. In all three cases we compare the impact of scheduling the same work
using a larger number of cgroups---120 functions---versus a baseline of
12 functions (i.e. \verb|12fn-CFS| and \verb|12fn-EEVDF|), which matches the number of hardware threads. The \verb|resctl-parallel| workload represents a case where each invocation requires two parallel worker threads that must both complete before the request can be returned. The \verb|resctl-mix| workload generates a mix of requests with multiple execution times, based on the  requirements of the Alibaba container-based serverless platform~\cite{Owl}: 30\% of requests at 10\,ms, 40\% at 100\,ms, and 30\% at 1{,}000\,ms.

Looking at the CDFs for \verb|resctl|, the performance of 120 functions under tuned CFS, RR, and tuned EEVDF shows that for simple workloads—where each invocation is mapped to a single thread with a known time slice in advance—tuning these schedulers with a sufficiently large default time slice can achieve results comparable to the improvements reported by \cfsllf. However, it is important to note that \cfsllf does not rely on modifying the default time slice of CFS; instead, it increases the total time allocated to a cgroup as a whole. Moreover, EEVDF performance actually improves under higher load: compare \verb|12fn-EEVDF| and \verb|120fn-EEVDF| for \verb|resctl| in Figure~\ref{fig:baselines-eevdf}. This is because EEVDF's  lag-based policy enforces low-level fairness at lower loads, while under higher loads it implicitly prefers task completion as task queuing leads to larger lags which are compensated for by longer execution times---highlighting the difficulty of tuning EEVDF to behave consistently~\cite{EEVDF-slice}.

Examining the performance for \verb|resctl-parallel|, we observe that tuning is less effective for CFS and has no impact at all for EEVDF. While tuned CFS and Linux RR improve median latency, their benefits remain limited compared to \cfsllf, since threads belonging to the same invocation do not necessarily complete simultaneously. \cfsllf shows the best performance in this case, with at least
50\% of requests completing with latencies comparable to the baseline, at
the cost of a slight increase in tail latency~\s{sec:host-evaluation:overhead}.

For \verb|resctl-mix|, \cfsllf again achieves the best overall latency distribution, albeit with a larger relative gap to the 12-function baseline for \verb|resctl-parallel| due to the heavier workload exceeding server capacity. Unlike tuned CFS, which primarily improves the latency of short invocations, \cfsllf effectively reduces latency across all request types. Linux RR performs worst of  all, underscoring the risks of relying on soft real-time policies.

\section{Related work}
\label{sec:related-works}

Prior work~\cite{WastedCores, Ipanema, Enoki} highlights the gap between kernel
scheduling policies and their practical implementations, particularly in the
context of multi-core, heterogeneous NUMA architectures. Our contribution is to
reduce the inherent costs of the scheduler itself under high workload
colocation, addressing the implementation challenges of managing per-cgroup
scheduling. Our evaluation of real-time group scheduling~\cite{Hierarchical-CBS},
\verb|RT_GROUP_SCHED| (disabled by default in the kernel), reveals even
more severe overhead issues due to the use of locks for enforcing CPU
time guarantees.

To the best of our knowledge, the only other attempt to address a
similar problem is a Linux patch proposal to flatten the cgroup hierarchy into a
single layer while calculating each cgroup’s share by compounding its share
across all hierarchy levels~\cite{cfs-flattened}. This appears to give
significant performance gains but it replaces significant sections of CFS code
and the authors acknowledge an additional starvation risk that can arise if a
large number of cgroups wake up simultaneously, not unlikely in a serverless
context~\cite{RunD, FaaSNet}.
\cfsllf reduces these overheads while preserving
group scheduling.

Some have investigated the efficiency of host scheduling for serverless
workloads~\cite{SFS,ALPS,Hermod} and identified the potential of approximating
SRTF given that invocations with short execution times dominate workloads.
\cfsllf leverages the broader assumption of
uneven resource usage during statistical multiplexing to improve workload colocation.
SFS~\cite{SFS} and ALPS~\cite{ALPS} address the challenge of
approximating the SRTF online to mitigate performance degradation when CPU
utilisation reaches between 70---90\%.
However, these are limited to ephemeral
functions in which a single process is instantiated per invocation and are not
directly applicable for kept-alive containers, which require application at the
granularity of corresponding cgroups~\s{sec:cfs-llf-implementation}.

\section{Conclusions}
\label{sec:conclusions}

We have shown that context switching overheads can become significant in serverless clusters due to reliance  on Linux group scheduling. Further, these overheads remain even under the new EEVDF policy as that preserves CFS' original goal of low-level fairness while continuing to depend on the group scheduling infrastructure. Our approach, \cfsllf, significantly reduces these overheads by using the Load Credit mechanism to prioritise task completion at the same time as ensuring effective use of all cores. This allows the same effective performance to be achieved for a realistic serverless workload on a 28\% smaller cluster.

Integrating \cfsllf into more recent kernels thus requires embedding Load Credit alongside the new heuristics that EEVDF introduced for cross-core task placement, which requires  consideration of orthogonal concerns such as cache locality and heterogeneous scheduling. Our preliminary evaluation on a high-end multi-socket server~\s{sec:host-evaluation} suggests that such porting is feasible, but a rigorous evaluation of the interactions between Load Credit and these other concerns is needed before such integration could be achieved.

% \break

\appendix
\section{Linux kernel patch for \cfsllf}\label{sec:patch}

\subsection{Implementation experience.}
Lozi et al.~\cite{WastedCores} identified bugs in CFS that violate its
work-conserving property, caused by the challenges of ensuring correctness in a
highly concurrent environment, especially in NUMA systems with hierarchical CPU
organisation and per-CPU core run queues. Our work addresses these issues and the
additional challenge of introducing \cfsllf at the right
cgroup granularity within the per-cgroup run queue architecture.

We implemented the new \cfsllf scheduler class iteratively, introducing multiple kernel flags to allow us to
assess the impact of adjusting various scheduler code paths including task
preemption, wakeups, time slices, and task placement. We use \cfsllf-static~\s{sec:cfs-llf-static} as a reference point to validate progress.
Recent techniques~\cite{Ipanema, Enoki, Plugsched} enable scheduler updates in a
running kernel which would significantly reduce development cycles. However,
they require addition of appropriate abstractions corresponding to per-cgroup
scheduling entities, which could then be abstracted through live-reloadable,
safe kernel modules. This could help avoid unprotected concurrent reads or
deadlocks, and other severe kernel bugs that cause the system to crash or hang.
However, while they do enable interaction with the core scheduler, they do not
expose the intricacies involved in managing per-cgroup run queues, which are
deeply embedded within CFS implementation. As such, we could not use them while implementing \cfsllf.

We ensure our implementation remains compatible with other CFS features except
for the proportional shares corresponding to \verb|cpu.shares| cgroup
properties. \cfsllf assumes that all cgroup weights provided by Kubernetes are
reset to the default weight of 1024 because CFS scales the load of scheduling
entities based on their relative weight, indirectly factoring in task priority
during load balancing. Adjusting these shares could conflict with the \cfsllf
policy or lead to inaccurate load estimations. This limitation could be addressed
in future work by providing a standalone implementation of the Load Credit
mechanism that is completely independent of PELT. Additionally, energy-awareness
features on heterogeneous CPU topologies, e.g.,~NUMA and Arm's big.LITTLE
architectures, could potentially override CFS’s load balancing heuristics for
task placement including our callback which incoporates the load metrics.
However, energy-aware placement~\cite{Nest} remains an orthogonal concern at
this point.

\subsection{Scheduler user space configuration}

Configuration is carried out in user space via the cgroup interface and kernel flags, and the design is readily portable to any cgroup-based serverless framework.

\textbf{Identifying function sandboxes}
Correctly applying the scheduling policy of \cfsllf requires identifying the
cgroups that correspond to serverless function sandboxes. We introduce a new
cgroup property to do so, \verb|cpu.latency_| \verb|awareness|. This is set to 0 by default,
and set to 1 for cgroups that represent a function. The implementation is based
on cgroup v2, Kubernetes' recommended cgroup API, requiring Linux kernel version
5.8 or later. In a Kubernetes-based serverless framework such as Knative, a
function is simply a pod and this interface can easily be configured
via:

\begin{verbatim}
echo 1 > /sys/fs/cgroup/cpu/kubepods.slice/
  kubepods-burstable.slice/*/cpu.latency_awareness
\end{verbatim}.

This makes the \cfsllf patch easily portable to serverless frameworks other than
Knative as most rely on cgroups to instantiate their function sandboxes whether
through processes, container runtimes or lightweight VMs.

\textbf{Tracking the Load Credit metric}\label{sec:cfs-llf-sched}
In vanilla CFS, the load metric is aggregated for every task group, maintained
across all the scheduling entities that belong to a cgroup across multiple cores
in the \verb|tg->load_avg| attribute. We extend the \verb|update_tg_load_avg|
function to calculate the exponential moving average (EMA) of this value into a
new attribute \verb|tg->load_avg_ema|, with the EMA window size exposed as a
kernel sysctl parameter, \verb|tg_load_avg_ema_window|, the impact of
which was explored earlier (Figure~\ref{fig:sensitivity-analysis}). It is
expressed in terms of number of scheduler ticks rather than absolute time due to
the cost and complexity of maintaining global time given the frequent update of
\verb|tg->load_avg| from run queues across multiple CPU cores.

\subsection{Scheduling code path modifications}
We adjust the key priority that drives CFS decisions in two code paths for
\one~single-core and \two~multi-core scheduling.

The first is the local code path that manages existing runnable tasks
(\verb|entity_before|) or tasks that become runnable on a single CPU core
(\verb|check_preempt_wakeup|). In vanilla CFS, runnable tasks are ordered in CFS
run queues in descending order by the \verb|vruntime| metric. When a task
becomes runnable, it can also preempt the current task if its \verb|vruntime|
significantly lags behind that of the current task. We retain this behaviour
across CFS run queues with the exception of scheduling entities that correspond
to a function sandbox. For these, we use the Load Credit metric of the
corresponding task group \verb|tg->load_avg_ema| instead of \verb|se->vruntime|.
Entities which are to be excluded from the default fair policy can be easily
identified thanks to the cgroup sandbox flag \verb|cpu.latency_awareness|.

The second is the global path responsible for deciding on which CPU to place a
task when it becomes runnable or when load balancing must be undertaken. This
code path starts from the \verb|select_task_rq_fair| hook in the scheduler and
implements CFS's work-conserving mechanism that prioritises allocation to idle
CPU cores. We extend these heuristics so that a task belonging to a lower load
credit function is placed to any CPU core currently running a task
belonging to a higher Load Credit function. The Load Credit of the currently
running task on each core is cached to enable quick comparison when deciding
when to place newly woken tasks. This is achieved with a callback
(\verb|cpu_has_higher_load_task|) invoked as CFS traverses the list of candidate
CPUs to run a woken up task.


\bibliographystyle{ACM-Reference-Format}
\bibliography{sosp25-isstaif}

\end{document}